\documentclass[twocolumn,showpacs,preprintnumbers,nofootinbib]{revtex4}%
\usepackage{amssymb}
\usepackage{amsmath}
\usepackage{graphicx}
\usepackage{epstopdf}
\usepackage{dcolumn}
\usepackage{bm}
\usepackage{amsfonts}%
\begin{document}
\title{Optical detection of paramagnetic defects in diamond using off-resonance
excitation of NV centers}
\author{Nir Alfasi}
\affiliation{Andrew and Erna Viterbi Department of Electrical Engineering, Technion, Haifa
32000 Israel}
\author{Sergei Masis}
\affiliation{Andrew and Erna Viterbi Department of Electrical Engineering, Technion, Haifa
32000 Israel}
\author{Oleg Shtempluck}
\affiliation{Andrew and Erna Viterbi Department of Electrical Engineering, Technion, Haifa
32000 Israel}
\author{Eyal Buks}
\affiliation{Andrew and Erna Viterbi Department of Electrical Engineering, Technion, Haifa
32000 Israel}
\date{\today }

\begin{abstract}
In this work we use fluorescence from nitrogen-vacancy defects in diamond to
detect and explore other paramagnetic defects in the diamond, such as P1 defects,
which are commonly undetectable through optical detection of magnetic
resonance in standard conditions. Our method does not require overlap between
the defects' resonances and therefore is applicable in a wide region of
magnetic fields and frequencies, as verified by excellent fit to theoretical
predictions. We propose a depolarization scheme of P1 defects to account for
the observed data. To verify our results, we perform cavity-based detection of
magnetic resonance and find a good agreement between the measured optically induced polarization and the value obtained
theoretically from rate equations. The findings in this work may open the way to
detection of paramagnetic defects outside of the diamond through the
photoluminesence of nitrogen-vacancy defects, which might be useful
for imaging in biology.

\end{abstract}
\maketitle



%It is always \today, today,
%but any date may be explicitly specified

%PACS, the Physics and Astronomy
%Classification Scheme.

\section{Introduction}

The nitrogen-vacancy (NV) defect in diamond consists of a substitutional
nitrogen atom (N) combined with a neighbor vacancy (V) \cite{Doherty_1}. In
its negatively-charged state the $\text{NV}^{-}$ defect has a spin triplet
ground state \cite{Doherty_205203} having relatively long coherence time
\cite{Balasubramanian_383}. The NV$^{-}$ spin state can be polarized via the
process of optically-induced spin polarization (OISP)
\cite{Robledo_025013,Redman_3420} and can be measured using the technique of
optical detection of magnetic resonance (ODMR)
\cite{Shin_124519,Chapman_190,Gruber_2012}. These properties facilitate a
variety of applications including magnetometry
\cite{Maze_644,Acosta_070801,Balasubramanian_648,Wolf_041001,Mamin_557,Pelliccione_700,Rondin_2279,Sushkov_197601}
and sensing \cite{Acosta_070801,Dolde_459,Balasubramanian_383,Jelezko_076401}.

ODMR of diamond samples having relatively low density of spin defects is
commonly described using a single-spin model, in which dipolar coupling
between different spins is disregarded. However, this approach becomes invalid
when the spin density is sufficiently high. In general, the strength of the
dipolar coupling can be characterized by the magnetic susceptibility $\chi$.
At the Larmor frequency, the magnetic susceptibility of a spin 1/2 ensemble
having a number density $n_{\mathrm{S}}$ is given by $\chi=i\left(
n_{\mathrm{S}}/n_{\mathrm{S}0}\right)  P_{z0}$, where $P_{z0}$ is the spin
polarization in steady state and the density $n_{\mathrm{S}0}$ is given by
$n_{\mathrm{S}0}=4T_{2}^{-1}/\hbar\gamma_{\mathrm{e}}^{2}\mu_{0}$,
where $T_{2}^{-1}$ is the transverse spin relaxation rate, $\gamma
_{\mathrm{e}}=2\pi\times28.03%
%TCIMACRO{\unit{GHz}}%
%BeginExpansion
\operatorname{GHz}%
%EndExpansion%
%TCIMACRO{\unit{T}}%
%BeginExpansion
\operatorname{T}%
%EndExpansion
^{-1}$ is the electron spin gyromagnetic ratio and $\mu_{0}$ is the
permeability of free space \cite{Abragam_Principles}. The transverse spin
relaxation rate of our sample $T_{2}^{-1}\simeq0.1%
%TCIMACRO{\unit{MHz}}%
%BeginExpansion
\operatorname{MHz}%
%EndExpansion
$ yields the value of $n_{\mathrm{S}0}\simeq10^{17}%
%TCIMACRO{\unit{cm}}%
%BeginExpansion
\operatorname{cm}%
%EndExpansion
^{-3}$, which is about 3 times smaller than the number density of NV$^{-}$
defects in our sample. Thus, when the polarization $\left\vert P_{z0}%
\right\vert $ becomes of order unity by applying OISP
\cite{Robledo_025013,Redman_3420}, $\chi$ cannot be treated as a small
parameter, and dipolar coupling cannot be disregarded.

In the current study we investigate diamond samples having relatively high
density of both NV$^{-}$ and nitrogen 14 substitution (P1) defects. The spin
density of our samples is not sufficiently high to allow access to the region
where macroscopic magnetic ordering occurs \cite{Ella_024414}, however it is
sufficiently high to make $\chi\gtrsim1$, i.e. to make effects originating
from dipolar coupling detectable. To further enhance such effects, the sample
is cooled down to a cryogenic temperature of $3.6%
%TCIMACRO{\unit{K}}%
%BeginExpansion
\operatorname{K}%
%EndExpansion
$ in order to decrease the rate of thermal polarization, which in turn makes OISP more efficient.

While only transitions between spin states with magnetic quantum numbers
$m_{\mathrm{s}}=\pm1$ and $m_{\mathrm{s}}=0$ of NV$^{-}$ are commonly visible
in the ODMR spectrum of diamond samples, our ODMR measurements reveal magnetic
resonances of P1 defects \cite{Wang_4135,Hall_10211,Purser_1802_09635}, as
well as resonances due to carbon 13 (C13) atoms, NV$^{-}$ $+1\rightarrow-1$
transitions and a resonance of unknown origin around $2~\mathrm{GHz}$ found at
magnetic fields near the NV$^{-}$ ground-state level anti-crossing (GSLAC).
The observation of nuclear and electronic spin transitions in P1 defects can
be attributed to driving-induced depolarization and a cross-relaxation process
\cite{Solomon_559,Belthangady_157601,Loretz_064413} between the $\text{NV}%
^{-}$ defects and the target spins under driving. These processes might be of
interest in the context of nuclear spins ensembles hyperpolarization
\cite{Takahashi_047601,Fischer_057601,Wang_1940}, with NV$^{-}$ defects
providing an optical path for readout and control.

To further explore dipolar coupling in our samples, the method of cavity-based
detection of magnetic resonance (CDMR) is employed \cite{Alfasi_063808}. As
discussed below, the results of the CDMR measurements provide further insight
into the underlying mechanism responsible for dipolar coupling in our samples.
In particular, these results clearly rule out the possibility that heating
plays an important role in this mechanism.

In addition, the CDMR measurements allow studying the process of OISP in
general, and the prospects of employing OISP for the generation of spin
population inversion, which, in-turn, may allow the realization of a
diamond-based maser. A maser is constructed by coupling an ensemble of
emitters to a microwave cavity and by applying external pumping for inducing
population inversion. A maser has a variety of applications, including
amplification of microwave signals, high-precision timing, optical to
microwave conversion, magnetometry and spectroscopy. The process of OISP can
be used for inducing population inversion in diamond when the externally
applied magnetic field is tuned to the region above the NV$^{-}$ GSLAC point.
This together with the relatively long coherence time
\cite{Balasubramanian_383} and long energy relaxation time \cite{Harrison_586}%
, makes the NV$^{-}$ defects in diamond useful for the construction of a
maser, as has been recently proposed \cite{Jin_9251} and
demonstrated experimentally \cite{Breeze_1710_07726}%
. In this work, we generate population inversion, though we do not reach the masing threshold due to anomalous
saturation of the polarization of NV$^{-}$ spins, which is consistent with previous
observations \cite{Loretz_064413,Drake_013011}.

\begin{figure}[pb]
\begin{center}
\includegraphics[
height=3.1142in,
width=3.4537in
]{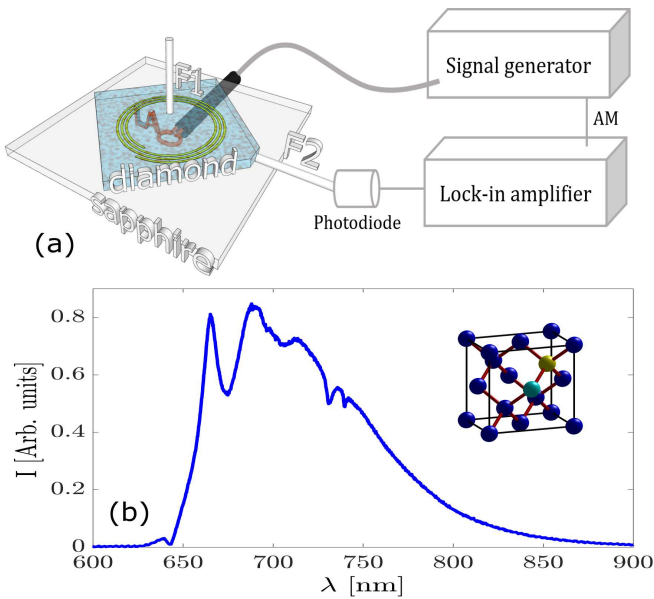}
\end{center}
\caption{Experimental setup. (a) Two multimode optical fibers F1 and F2 are glued to the
diamond, which is attached to a sapphire wafer supporting a spiral-shaped
superconducting resonator. A loop antenna (LA) transmits both input and output microwave
signals. (b) PL intensity $I$ as a function of optical wavelength $\lambda$.}%
\label{Fig_Setup}%
\end{figure}
%EndExpansion

\begin{figure*}[tb]
\begin{center}
\includegraphics[width=1.6\columnwidth]{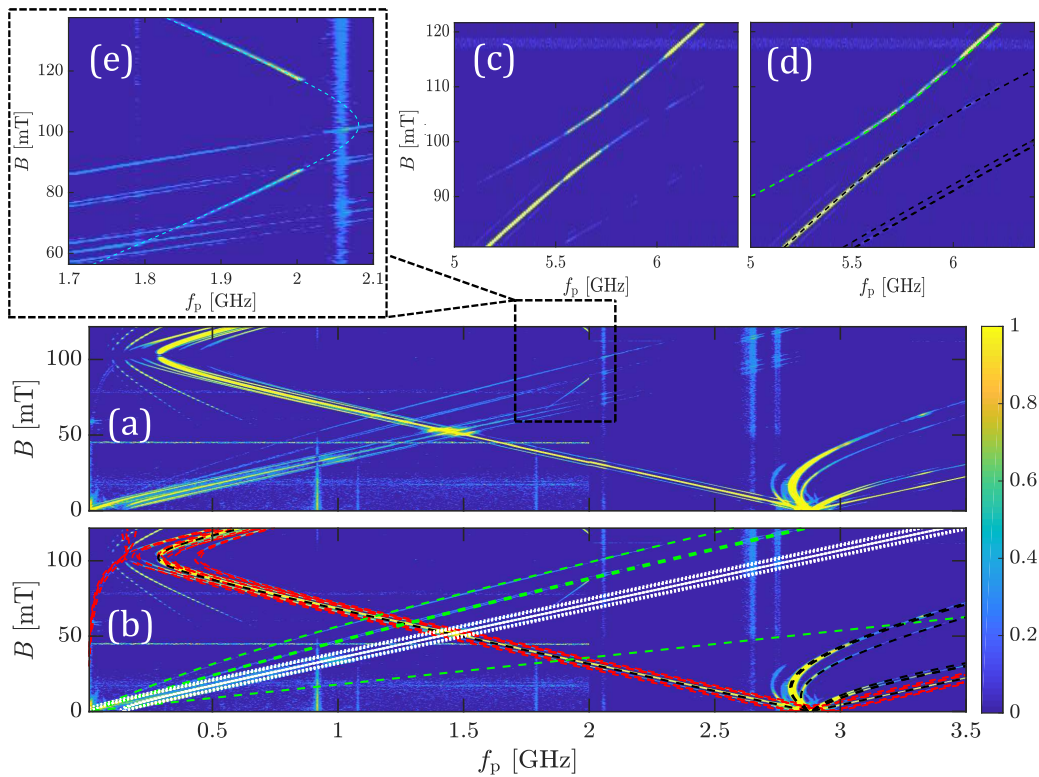}
\end{center}
\caption{ODMR. (a) Color coded plot of the normalized ODMR signal as a
function of frequency $f_{\mathrm{p}}$ and magnetic field $B$. (b) The
overlaid black (green) dashed lines represent NV$^{-}$ ground state $\pm1$ to
$0$ ($+1$ to $-1$) transitions and are calculated by numerically diagonalizing
the NV$^{-}$ ground state spin triplet Hamiltonian $\mathcal{H}_{\mathrm{NV}%
^{-}}$ (\ref{H NV- w/o 13C}). The red dashed lines are obtained by adding the
hyperfine interaction term $\mathcal{V}_{\mathrm{NV}^{-},\mathrm{C}}$
(\ref{V_NV,13C}) to the Hamiltonian $\mathcal{H}_{\mathrm{NV}^{-}}$
(\ref{H NV- w/o 13C}). The white dotted lines represent P1 electronic-like
transitions, calculated using the Hamiltonian $\mathcal{H}_{\mathrm{P}1}$
(\ref{H_P1}). Higher oreder hyperbolas corresponding to multi-photon
resonances are also observed and are discussed in \cite{Masis2019}. The
following values are assumed in the calculations: $D%
=2\pi\times2.88\operatorname{GHz}$, $E_{\mathrm{NV}^{-}}=2\pi\times
10\operatorname{MHz}$ \cite{Ovartchaiyapong_1403_4173,MacQuarrie_227602},
$A_{\mathrm{C},\parallel}=2\pi\times199.7\operatorname{MHz}$, $A_{\mathrm{C}%
,\perp}=2\pi\times120.3\operatorname{MHz}$
\cite{Felton_075203,Mizuochi_041201,Simanovskaia_224106,Kamp_045204,Loubser_1201}%
, $\gamma_{\mathrm{n}}=2\pi\times3.0766\operatorname{MHz}\operatorname{T}%
^{-1}$, $Q_{\mathrm{P}1}=-2\pi\times3.97\operatorname{MHz}$, $A_{\mathrm{P}%
1,\parallel}=2\pi\times114\operatorname{MHz}$ and $A_{\mathrm{P}1,\perp}%
=2\pi\times81.3\operatorname{MHz}$ \cite{Cox_551}. (c) ODMR in the high
frequency regime showing the ground state level anticrossing (GSLAC). (d) See
(b). (e) Resonance lines of unknown origin. Cyan dotted line represents fit to Eq.~\eqref{eq:unknownres}.}%
\label{Fig_ODMR_S1}%
\end{figure*}
%EndExpansion

\section{Experimental Setup}

The experimental setup is schematically depicted in Fig.~\ref{Fig_Setup}.
Defects in a [110] type Ib diamond are created using $2.8%
%TCIMACRO{\unit{MeV}}%
%BeginExpansion
\operatorname{MeV}%
%EndExpansion
$ electron irradiation with a dose of approximately $8\times10^{18}%
~\mathrm{e/cm^{2}}$, followed by annealing at $900^{\circ}\mathrm{C}$ for 2
hours and acid cleaning, resulting in the formation of NV$^{-}$ defects with
density of $n_\mathrm{S}\simeq3.25\times10^{17}%
%TCIMACRO{\unit{cm}}%
%BeginExpansion
\operatorname{cm}%
%EndExpansion
^{-3}$ \cite{Farfurnik_123101}.

Two multimode optical fibers are glued to the irradiated diamond using optical adhesive.
The one labeled as F1 in Fig.~\ref{Fig_Setup}(a) is used for delivering laser
light having wavelength of $532%
%TCIMACRO{\unit{nm}}%
%BeginExpansion
\operatorname{nm}%
%EndExpansion
$ to the sample, and the one labeled as F2 transmits the photoluminescence
(PL) emitted from the sample. The diamond is attached to a sapphire wafer supporting a spiral shaped
superconducting microwave cavity (resonator) \cite{Maleeva_474,Maleeva_064910}. The spiral is made of $180~\mathrm{nm}$ thick niobium, has 7-turns, inner diameter of $0.7~\mathrm{mm}$ and width and spacing of $40~\mathrm{\mu m}$. In the first part of this work, the delivered light intensity is relatively high and consequently the resonator is
heated above its critical temperature and is effectively disabled, while in
the second part, the light intensity is reduced and the resonator may be used
as a superconducting cavity. A three-loop antenna (LA) is placed in the
vicinity of the diamond. An amplitude modulated signal at frequency
$f_{\mathrm{p}}$ is injected into the LA, and the modulation signal serves as
a reference for a lock-in amplifier connected to a photodiode collecting the
filtered PL ($605-700%
%TCIMACRO{\unit{nm}}%
%BeginExpansion
\operatorname{nm}%
%EndExpansion
$). A superconducting solenoid magnet is used for applying magnetic field
$\mathbf{B}$. The setup is cooled down to a base temperature of $3.6%
%TCIMACRO{\unit{K}}%
%BeginExpansion
\operatorname{K}%
%EndExpansion
$ to enhance hyperpolarization of non-NV impurities \cite{Ardenkjaer_10158}.

The spatial orientation of the sample with respect to the externally applied
magnetic field is characterized by a unit vector $\mathbf{\hat{n}%
}_{\mathrm{MA}}=\left(  \sin\theta_{\mathrm{MA}}\cos\varphi_{\mathrm{MA}}%
,\sin\theta_{\mathrm{MA}}\sin\varphi_{\mathrm{MA}},\cos\theta_{\mathrm{MA}%
}\right)  $. In this work, we use several orientations to confirm our
theoretical predictions. The angles for the data presented in
Fig.~\ref{Fig_ODMR_S1} and Fig.~\ref{Fig_P1_NV} are $\theta_{\mathrm{MA}%
}=-4^{\circ}$ and $\varphi_{\mathrm{MA}}=95^{\circ}$, in
Fig.~\ref{Fig_ODMR_S2} $\theta_{\mathrm{MA}}=-12.3^{\circ}$ and $\varphi
_{\mathrm{MA}}=95^{\circ}$, in Fig.~\ref{Fig_ODMR_S3} $\theta_{\mathrm{MA}%
}=-9.5^{\circ}$ and $\varphi_{\mathrm{MA}}=94.5^{\circ}$, whereas the angles
used for the CDMR measurements in Fig.~\ref{fig:gamma2} are $\theta
_{\mathrm{MA}}=-1.85^{\circ}$ and $\varphi_{\mathrm{MA}}=97^{\circ}$.

\section{NV$^{-}$}

The NV$^{-}$ ground state spin triplet Hamiltonian is given by
\cite{Ovartchaiyapong_1403_4173,MacQuarrie_227602}%
\begin{align}
\frac{\mathcal{H}_{\mathrm{NV}^{-}}}{\hbar}  &  =D%
S_{1,z}^{2}-\gamma_{\mathrm{e}}\mathbf{B}\cdot\mathbf{S}_{1}+\frac
{E_{\mathrm{NV}^{-}}\left(  S_{1,+}^{2}+S_{1,-}^{2}\right)  }{2}\;,\nonumber\\
&  \label{H NV- w/o 13C}%
\end{align}
where $D$ is the zero field splitting induced by spin-spin
interaction, $\mathbf{S}_{1}=\left(  S_{1,x},S_{1,y},S_{1,z}\right)  $ is a
vector electronic spin $1$ operator, $\mathbf{B}$ is an externally applied
magnetic field, $E_{\mathrm{NV}^{-}}$ is the strain-induced splitting, the
raising $S_{1,+}$ and lowering $S_{1,-}$ operators are defined by $S_{1,\pm
}=S_{1,x}\pm iS_{1,y}$ and the NV$^{-}$ axis is parallel to the $z$ axis.

\begin{figure}[ptb]
\begin{center}
\includegraphics[
width=0.95\columnwidth
]{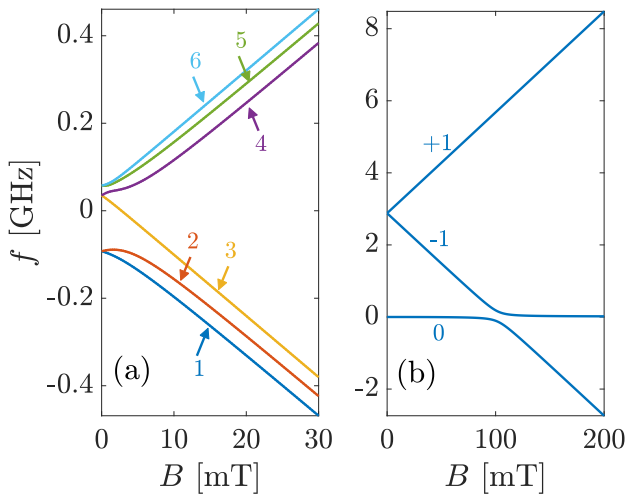}
\end{center}
\caption{Energies of P1 (a) [see Eq. (\ref{H_P1})] and NV$^{-}$ (b) [see Eq.
(\ref{H NV- w/o 13C})] states (in frequency units) vs. magnetic field for the case where the angles
between the axis and the magnetic field are given by $\theta_{\mathrm{MA}%
}=-4^{\circ}$ and $\varphi_{\mathrm{MA}}=95^{\circ}$.}%
\label{Fig_P1_NV}%
\end{figure}

The color-coded plot shown in Fig.~\ref{Fig_ODMR_S1}(a) represents the
normalized lock-in signal as a function of the frequency of the applied
microwave signal $f_{\mathrm{p}}$ and the externally applied stationary
magnetic field $B$. In a single crystal diamond the NV defects have four
different possible orientations. The two frequencies per orientation
corresponding to the transitions between the spin state with magnetic quantum
number $0$ and the spin state with magnetic quantum number $\pm1$ are
calculated by diagonalizing the Hamiltonian~(\ref{H NV- w/o 13C}) [see Fig.~\ref{Fig_P1_NV}(b)], and are
shown in by the black dashed lines in Fig.~\ref{Fig_ODMR_S1}(b). A fitting
procedure allows determining the sample orientation (characterized by the
angles $\theta_{\mathrm{MA}}$ and $\varphi_{\mathrm{MA}}$) with respect to the
externally applied magnetic field. Additional resonance frequency per NV$^{-}$
orientation is seen in the ODMR spectra and is marked by green dashed lines in
Fig.~\ref{Fig_ODMR_S1}(b,d). It corresponds to the transition between states
with magnetic quantum numbers $+1$ and $-1$. When the NV$^{-}$ vector and the
magnetic field are not strictly parallel, or near the GSLAC, these states
acquire a non negligible component of $0$ spin state, and hence the transition between them becomes
measurable with ODMR \cite{Doherty_1}. At some combinations of the magnetic
field orientation and magnitude, these transitions might be useful for
magnetometry with higher responsivity than the one obtained with the more commonly used $0\rightarrow\pm1$ transitions, due to the two-times larger pre-factor of $2\gamma_{\mathrm{e}}$. In order to verify our results, we repeat the same procedure using different orientation of the sample. For both cases excellent fit is obtained (see Figs.~\ref{Fig_ODMR_S1} and \ref{Fig_ODMR_S2}).

\section{Carbon 13}

The hyperfine interaction between an NV$^{-}$ defect and the nearest-neighbor
carbon-13 atom has been studied in
\cite{Felton_075203,Mizuochi_041201,Simanovskaia_224106,Kamp_045204,Loubser_1201}%
. To account for this interaction, a coupling term $\mathcal{V}_{\mathrm{NV}%
^{-},\mathrm{C}}$ given by \cite{Slichter_Principles,Cox_551}%
\begin{equation}
\frac{\mathcal{V}_{\mathrm{NV}^{-},\mathrm{C}}}{\hbar}=\mathbf{S}%
_{1}R_{\mathrm{C}}^{-1}\mathcal{A}_{\mathrm{C}}R_{\mathrm{C}}\mathbf{I}%
_{1/2}^{\mathrm{T}}\;, \label{V_NV,13C}%
\end{equation}
is added to the Hamiltonian $\mathcal{H}_{\mathrm{NV}^{-}}$
(\ref{H NV- w/o 13C}), where $\mathbf{I}_{1/2}$ is a vector nuclear spin $1/2$
operator, the matrix $\mathcal{A}_{\mathrm{C}}$ is given by $\mathcal{A}%
_{\mathrm{C}}=\operatorname{diag}\left(  A_{\mathrm{C},\perp},A_{\mathrm{C}%
,\perp},A_{\mathrm{C},\parallel}\right)  $, where $A_{\mathrm{C},\parallel}$
and $A_{\mathrm{C},\perp}$\ are respectively the longitudinal and transverse
hyperfine coupling parameters. The matrix $R_{\mathrm{C}}$ is a rotation
matrix satisfying $R_{\mathrm{C}}\mathbf{\hat{n}}_{\mathrm{C}}=\mathbf{\hat
{z}}$, where $\mathbf{\hat{n}}_{\mathrm{C}}$ is a unit vector parallel to the
vacancy-carbon axis and $\mathbf{\hat{z}}$ is a unit vector in the $z$
direction. The overlaid red dashed lines in Figs.~\ref{Fig_ODMR_S1}(b) and \ref{Fig_ODMR_S2}(b) are
calculated by numerically diagonalizing the NV$^{-}$ Hamiltonian
(\ref{H NV- w/o 13C}) with the added hyperfine coupling term (\ref{V_NV,13C}).

\section{P1}

A nitrogen 14 (nuclear spin 1) substitution defect (P1) in diamond
\cite{Smith_1546,Cook_99,Loubser_1201} has four locally stable configurations.
In each configuration a static Jahn-Teller distortion \cite{Smith_1546}
occurs, and an unpaired electron is shared by the nitrogen atom and by one of
the four neighboring carbon atoms, which are positioned along one of the
lattice directions $\left\langle 111\right\rangle $
\cite{Takahashi_047601,Hanson_087601,Hanson_352,Wang_4135,Broadway_1607_04006,Shim_1307_0257,Smeltzer_025021,Shin_205202,Clevenson_021401,Schuster_140501}%
.

\begin{figure*}[ptb]
\begin{center}
\includegraphics[width=1.6\columnwidth]{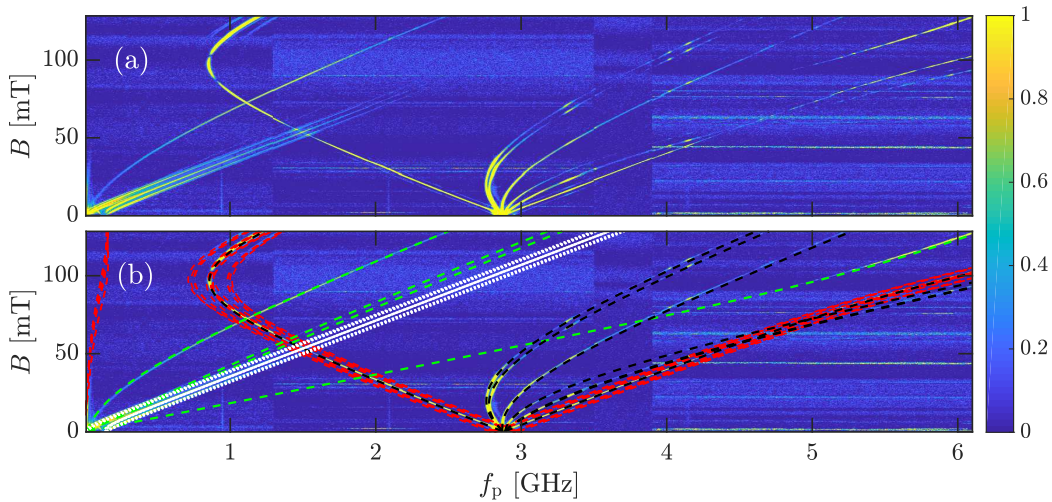}
\end{center}
\caption{Rotated configuration. (a) ODMR data obtained by rotating the sample
by about $10^{\circ}$ with respect to the configuration used for generating
the data shown in Fig.~\ref{Fig_ODMR_S1}(a). (b) Theoretical calculation of
resonance lines due to NV$^{-}$ $\pm1\rightarrow0$ transitions (black),
NV$^{-}$ $+1\rightarrow-1$ transitions (green), carbon 13 (red) and P1 (white)
defects. All lines were obtained by modifying only the diamond rotation
angles.}%
\label{Fig_ODMR_S2}%
\end{figure*}

The overlaid white dotted lines in Figs.~\ref{Fig_ODMR_S1}(b) and \ref{Fig_ODMR_S2}(b) are calculated by
numerically diagonalizing the P1 spin Hamiltonian $\mathcal{H}_{\mathrm{P}1}$
[see Fig.~\ref{Fig_P1_NV}(a)], which is given by \cite{Cox_551}%
\begin{align}
\frac{\mathcal{H}_{\mathrm{P}1}}{\hbar}  &  =\gamma_{\mathrm{e}}%
BS_{1/2,z}+\gamma_{\mathrm{n}}BI_{1,z}+Q_{\mathrm{P}1}I_{1,z}^{2}\nonumber\\
&  +\mathbf{S}_{1/2}R_{\mathrm{P}}^{-1}\mathcal{A}_{\mathrm{P}1}R_{\mathrm{P}%
}\mathbf{I}_{1}^{\mathrm{T}}\;,\nonumber\\
&  \label{H_P1}%
\end{align}
where $\gamma_{\mathrm{n}}$ is the nitrogen 14 nuclear gyromagnetic ratio,
$Q_{\mathrm{P}1}$ is the nitrogen 14 quadrupole coupling, the matrix
$\mathcal{A}_{\mathrm{P}1}$ is given by $\mathcal{A}_{\mathrm{P}%
1}=\operatorname{diag}\left(  A_{\mathrm{P}1,\perp},A_{\mathrm{P}1,\perp
},A_{\mathrm{P}1,\parallel}\right)  $, where $A_{\mathrm{P}1,\parallel}$ and
$A_{\mathrm{P}1,\perp}$\ are respectively the longitudinal and transverse
hyperfine coupling parameters, $\mathbf{S}_{1/2}$ is an electronic spin $1/2$
vector operator and $\mathbf{I}_{1}$ is a nuclear spin $1$ vector operator.
The matrix $R_{\mathrm{P}}$ is a rotation matrix satisfying $R_{\mathrm{P}%
}\mathbf{\hat{n}}_{\mathrm{P}}=\mathbf{\hat{z}}$, where $\mathbf{\hat{n}%
}_{\mathrm{P}}$ is a unit vector parallel to the P1 axis.

\begin{figure*}[ptb]
\begin{center}
\includegraphics[width=1.6\columnwidth]{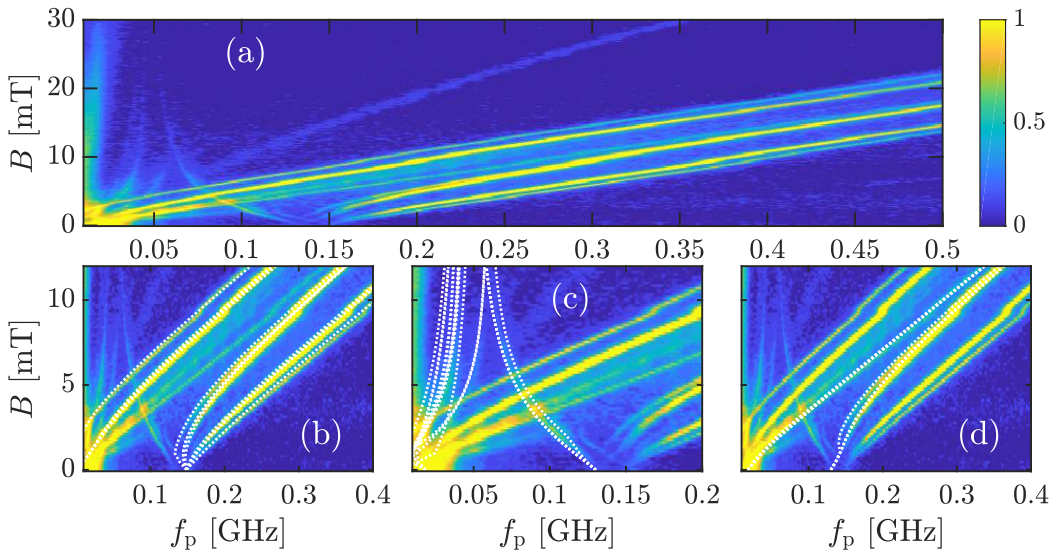}
\end{center}
\caption{Normalized ODMR in the region of relatively small magnetic field. The overlaid
white dotted lines represent (b) electronic-like, (c) nuclear-like and (d)
mixed transitions in P1.}%
\label{Fig_ODMR_S3}%
\end{figure*}
%EndExpansion

Higher resolution ODMR measurements of the region of small magnetic field are
shown in Fig.~\ref{Fig_ODMR_S3}(a). The overlaid white dotted lines in (b),
(c) and (d) represent P1 resonances. The $6$ states per axis orientation are
numbered in Fig.~\ref{Fig_P1_NV}(a). Electronic-like transitions corresponding
to the pairs $\left(  6,1\right)  $, $\left(  5,2\right)  $ and $\left(
4,3\right)  $ are seen in Fig.~\ref{Fig_ODMR_S3}(b), nuclear-like transitions
corresponding to the pairs $\left(  6,5\right)  $, $\left(  5,4\right)  $,
$\left(  3,2\right)  $ and $\left(  2,1\right)  $ in Fig.~\ref{Fig_ODMR_S3}(c)
and two mixed transitions corresponding to the pairs $\left(  6,3\right)  $
and $\left(  5,3\right)  $ in Fig.~\ref{Fig_ODMR_S3}(d). To our knowledge,
this is the first reported observation of P1 nuclear-like transitions through PL of NV
defects in diamond.

The visibility of P1 nuclear and electronic spin transitions in the ODMR data could be
explained by resonant microwave heating, however this hypothesis contradicts
CDMR measurements, as will be discussed below in Sec.~\ref{sec:CDMR}.
Alternatively, we attribute the visibility of P1 resonances in the ODMR data
to a reverse-hyperpolarization process. In general, large imbalance in spin
polarization between P1 and NV$^{-}$ defects can be induced by applying OISP,
which significantly enhances spin polarization of NV$^{-}$ defects. In the
absent of any coupling between P1 and NV$^{-}$ defects, the spin polarization
of P1 defects is expected to be unaffected by the optical pumping, and remains
at the value corresponding to thermal equilibrium. However, finite coupling is
expected to give rise to optically-induced enhancement in the P1 spin
polarization. Such enhancement is observed in our CDMR measurements and is
discussed in Sec.~\ref{sec:CDMR} below. On the other hand, the opposite effect
of P1 depolarization can be induced by applying radiofrequency (RF) signal at a frequency close
to a P1 transition resonance. Such P1 driving-induced depolarization, in-turn, is expected to
give rise to NV$^{-}$ depolarization, due to the coupling between the two
ensembles. The observed change in the ODMR signal is attributed according to
this mechanism to such an NV$^{-}$ depolarization. Rate equations for this
reverse-hyperpolarization process are given in Appendix A.

\section{Unidentified resonance}

While we have been able to identify most resonance lines in the ODMR plot, an
unidentified pronounced feature appears when better alignment is achieved [see
Fig.~\ref{Fig_ODMR_S1}(e)]. The observation that this feature is symmetrical around
$\sim102~\mathrm{mT}$ (GSLAC of NV$^{-}$) suggests that it originates from a coupling between the NV$^{-}$ defects having axis nearly parallel to \textbf{B} and a resonance having a magnetic field independent frequency. However, we could not find any
evidence for such resonance in our measurements (both optical and microwave
reflection). Interestingly, the frequency $F_\mathrm{f}$ of this resonance line may be fitted by using the following relation [see cyan dashed line in Fig.~\ref{Fig_ODMR_S1}(e)]
\begin{equation}
F_\mathrm{f}=F_\mathrm{s}-\frac{F_\mathrm{NV^-}}{3},
\label{eq:unknownres}
\end{equation}
where $F_\mathrm{s}=2.169~\mathrm{GHz}$ and $F_\mathrm{NV^-}$ is the \textbf{B}-dependent resonance frequency of the $-1\rightarrow0$ transition of the NV$^-$ defects having axis nearly parallel to \textbf{B} [see black dashed lines in Fig.~\ref{Fig_ODMR_S1}(b)]. In a different cooling cycle and alignment, we find Eq.~\eqref{eq:unknownres} to fit the data well with $F_\mathrm{s}=2.172~\mathrm{GHz}$. 

\section{CDMR measurements}

\label{sec:CDMR}

\begin{figure*}[ptb]
\begin{center}
\includegraphics[width=1.6\columnwidth]{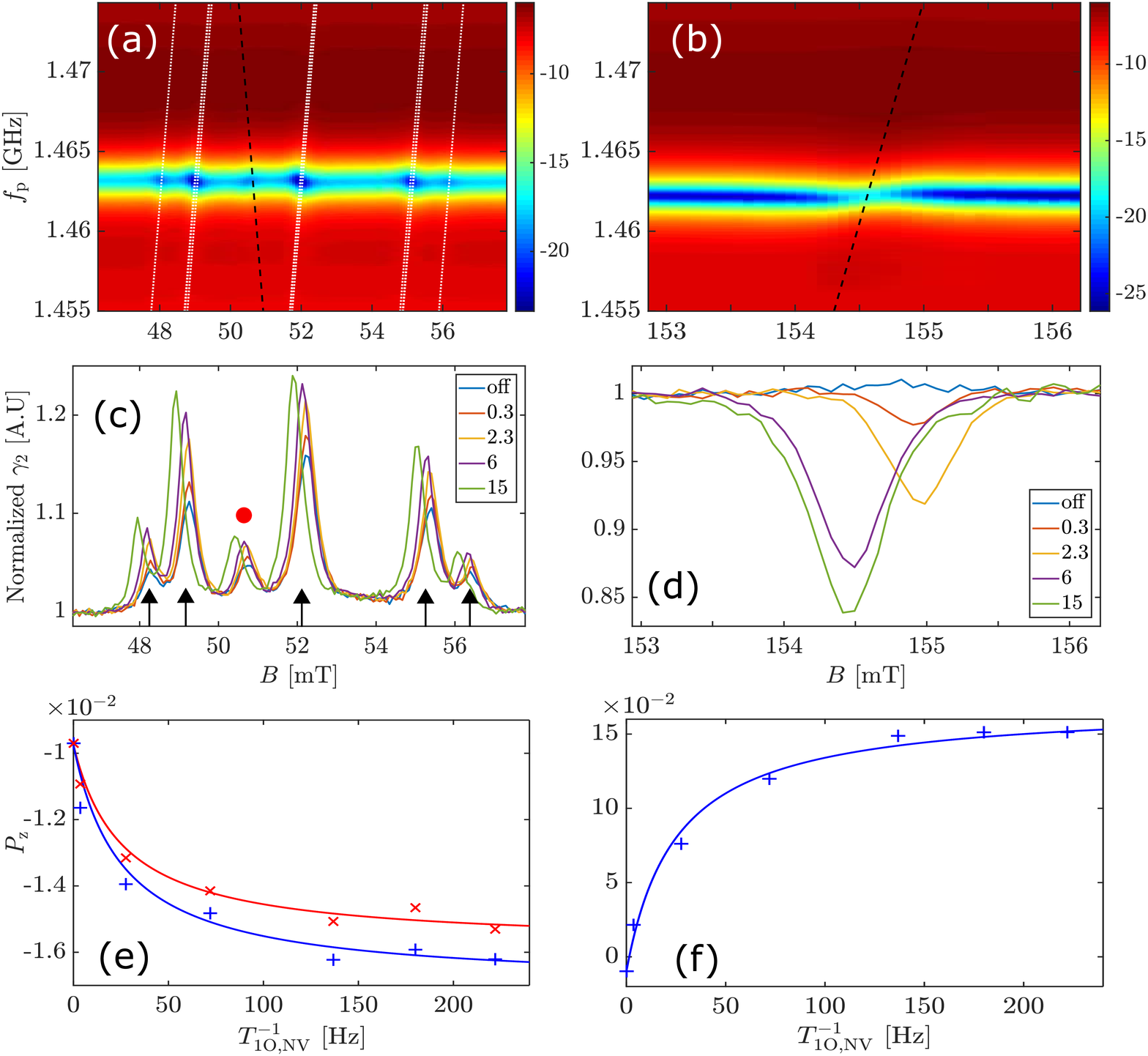}
\end{center}
\caption{Cavity mode reflectivity showing (a) five P1 resonances and one
NV$^{-}$ resonance and (b) one NV$^{-}$ resonance in the over-coupled
resonator case. Fit to theory is shown using black dashed lines for
NV$^{-}$ $0\rightarrow-1$ transition [see Eq.~(\ref{H NV- w/o 13C})] and white
dotted lines for P1 defects [see Eq.~(\ref{H_P1})]. (c)-(d) Change in normalized
$\gamma_{2}$ with respect to the magnetic field for various laser powers
(legends inside figures indicate values of laser intensity $I_{\mathrm{L}}$ in
units of $\operatorname{mW}\operatorname{mm}^{-2}$). Resonances due to P1
defects are marked by black arrows, while resonance due to NV$^{-}$
$0\rightarrow-1$ transition is marked by a red circle. (e)-(f) Polarization $P_{z}$ as a function of OISP rate $T_{1\mathrm{O}%
}^{-1}$. Markers (blue pluses for NV$^{-}$, red crosses for P1) denote data
extracted from (c)-(d), whereas solid lines (blue for NV$^{-}$, red for P1) represent the steady state numerical solution of the rate equations
(\ref{P_z,P1 eom}) and (\ref{P_z,NV eom}). The parameters used for the
calculations are $C_{\mathrm{O}}=1.5$, $P_{z\mathrm{T}}=-9.7\times10^{-3}$,
$T_{1\mathrm{T},\mathrm{NV}}^{-1}=25~\operatorname{Hz}$, $T_{\mathrm{I}%
,\mathrm{NV}}^{-1}=5~\operatorname{Hz}$, $T_{1\mathrm{T},\mathrm{P1}}%
^{-1}=8~\operatorname{Hz}$, $T_{\mathrm{I},\mathrm{P1}}^{-1}%
=40~\operatorname{Hz}$ and $T_{\mathrm{d},\mathrm{P1}}^{-1}=0$.}%
\label{fig:gamma2}%
\end{figure*}

In this section, we reduce the injected light intensity to avoid
heating of the superconducting spiral resonator located below our diamond sample, so
the resonator regains its superconductive properties and may be used as a
cavity. We employ the method of CDMR as discussed in \cite{Alfasi_063808}. We
apply reflection measurements using a network analyzer to extract the reflection coefficient $S_{11}$,
which is given by \cite{Abdo2006}
\begin{equation}
S_{11}=\frac{i(\omega_{\mathrm{p}}-\omega_{0})-(\gamma_{2}-\gamma_{1}%
)}{i(\omega_{\mathrm{p}}-\omega_{0})-(\gamma_{2}+\gamma_{1})}\;,
\end{equation}
where $\omega_{0}\simeq2\pi\times1.464~\mathrm{GHz}$ is the cavity angular
resonance frequency, $\omega_{\mathrm{p}}$ is the angular frequency of the
injected signal and $\gamma_{1}$ and $\gamma_{2}$ are coupling coefficients
associated with the antenna-resonator and the resonator-reservoir
(dissipation) couplings, respectively. Note that the cavity resonance
frequency is set to the crossing point between NV$^{-}$ $-1\rightarrow0$ and
P1 electronic-like transitions. We fit our data and find $\gamma_{1}=1.15~%
%TCIMACRO{\unit{MHz}}%
%BeginExpansion
\operatorname{MHz}%
%EndExpansion
$ and $\gamma_{2}=0.62~\operatorname{MHz}$ (laser off), leading to a Q-factor of
$\sim2500$, and by assuming $\gamma_{1}$ is independent of laser power, we
find $\gamma_{2}$ for various laser powers. Note that $\gamma_{1}>\gamma_{2}$,
i.e. the resonator is over-coupled (away from spin resonances).
Figure~\ref{fig:gamma2}(a) shows typical CDMR data
with six resonances - five corresponding to P1 transitions and one to the
NV$^{-}$ $-1\rightarrow0$ transition, whereas Fig.~\ref{fig:gamma2}(b) is for
magnetic field above the GSLAC, where the ground state of the NV$^{-}$ triplet
is the state having $m_{\mathrm{s}}=-1$ and the transition $0\rightarrow-1$ is seen. In the
first case, we expect the coupling between the NV$^{-}$ defects and the
superconducting resonator to \textit{increase} $\gamma_{2}$ due to increased
losses. The opposite behavior is expected in the second case, since the state $m_{\mathrm{s}}=0$, which is optically populated via OISP, is no longer the ground state
of the system, and consequently some population inversion is expected, resulting in a
\textit{decrease} in $\gamma_{2}$. These phenomena are observed here and are
shown in Figs.~\ref{fig:gamma2}(c)-(d) for various laser intensities $I_{\mathrm{L}}$. For both cases, the process of optically-induced change in $\gamma_{2}$ exhibits a saturation for $I_{\mathrm{L}}\sim10-15~%
%TCIMACRO{\unit{mW}}%
%BeginExpansion
\operatorname{mW}%
%EndExpansion%
%TCIMACRO{\unit{mm}}%
%BeginExpansion
\operatorname{mm}%
%EndExpansion
^{-2}$, which is similar to the anomalous saturation reported in
\cite{Loretz_064413,Drake_013011}.

In an attempt to increase the loaded $Q$-factor, we move the antenna
further away from the diamond-resonator sample. The reduced coupling to the antenna makes the resonator under-coupled ($\gamma_{1}%
<\gamma_{2}$) with $\gamma_{1}=0.27~%
%TCIMACRO{\unit{MHz}}%
%BeginExpansion
\operatorname{MHz}%
%EndExpansion
$. We repeat the same measurements and similar results are obtained.

The expected change in normalized cavity damping, which is denoted by
$\vartheta=\gamma_{\mathrm{s}}/\gamma_{\mathrm{2}}$, where $\gamma
_{\mathrm{s}}$ is the spin-induced change in the cavity damping rate, can be
estimated using Eqs. (4) and (8) of Ref. \cite{Alfasi_063808}%
\begin{equation}
\vartheta=\frac{\kappa}{1+\Delta^{2}T_{2}^{2}}\frac{1+\frac{T_{1\mathrm{O}%
}^{-1}}{T_{1\mathrm{T}}^{-1}}\frac{P_{z\mathrm{SO}}\left(  I_{\mathrm{L}%
}\right)  }{P_{z\mathrm{ST}}}}{1+\frac{T_{1\mathrm{O}}^{-1}}{T_{1\mathrm{T}%
}^{-1}}}{},\label{vartheta}%
\end{equation}
where $\kappa$ is the cooperativity parameter characterizing the coupling
between the spins and the cavity mode, $\Delta$ is the frequency detuning
between the cavity mode frequency and the spin's transition frequency and
$T_{1\mathrm{O}}^{-1}$ ($T_{1\mathrm{T}}^{-1}$) is the rate of OISP (thermal
relaxation). In steady state and when $T_{1\mathrm{T}}^{-1}\gg T_{1\mathrm{O}%
}^{-1}$ (i.e. when OISP is negligibly small) the coefficient $P_{z\mathrm{ST}%
}$ is the value of spin polarization $P_{z}$ in thermal equilibrium. In the
opposite limit of $T_{1\mathrm{O}}^{-1}\gg T_{1\mathrm{T}}^{-1}$ (i.e. when
thermal relaxation is negligibly small) the coefficient $P_{z\mathrm{SO}}$ is
the value of $P_{z}$ in steady state. The rate $T_{1\mathrm{O}}^{-1}$ of OISP
is expressed as $T_{1\mathrm{O}}^{-1}=C_{\mathrm{O}}\gamma_{\mathrm{O}}$,
where $C_{\mathrm{O}}$ is a dimensionless parameter characterizing the
absorption efficiency and where $\gamma_{\mathrm{O}}=I_{\mathrm{L}}%
\sigma\lambda_{\mathrm{L}}/hc$ is the rate of optical absorption,
$I_{\mathrm{L}}$ is the laser intensity, $\sigma=3\times10^{-17}%
%TCIMACRO{\unit{cm}}%
%BeginExpansion
\operatorname{cm}%
%EndExpansion
^{2}$ \cite{Wee_9379} is the optical cross section, $h$ is the Plank's
constant and $c$ is the speed of light in vacuum. Note that Eq.
(\ref{vartheta}) is obtained by assuming that all contributing spins share the
same coupling to the cavity mode and to the laser light and that nonlinearity
in the response can be disregarded.

The spin polarization $P_{z}$ as a function of OISP rate $T_{1\mathrm{O}}^{-1}$ can be extracted from the data shown in Fig.~\ref{fig:gamma2} using Eq.~(\ref{vartheta}). Results of the extraction procedure are presented in Fig.~\ref{fig:gamma2}(e)-(f) using
blue pluses (red crosses) markers for NV$^{-}$ (P1) for magnetic fields below
[Fig.~\ref{fig:gamma2}(e)] and above [Fig.~\ref{fig:gamma2}(f)] the NV$^{-}$
GSLAC. It can be seen from Fig.~\ref{fig:gamma2}(e) that when the NV$^{-}$ and P1 resonances coincide,
polarization of both defects is of the same order of magnitude and OISP is
$\sim1.6$ times larger when compared to thermal polarization alone (laser
off). The rate equations derived in appendix A are used to calculate the expected dependency of $P_{z}$ on $T_{1\mathrm{O}}^{-1}$. Solid lines in Fig.~\ref{fig:gamma2}(e)-(f) are generated by solving
Eqs.~\eqref{P_z,P1 eom} and \eqref{P_z,NV eom} (the parameters used for the
fit are given in the caption of Fig.~\ref{fig:gamma2}). The comparison
between data and theory yields a good agreement.

In order to gain a further insight, we revisit CDMR data acquired with a different,
yet similar, setup as described in \cite{Alfasi_063808}, where NV$^{-}$ and P1
resonances do not coincide near the cavity frequency. Same analysis as
described above is performed and OISP is found to be about $5$ times larger for
NV$^{-}$ when compared to thermal polarization alone, but only $~1.15$ larger for P1
defects. These results suggest that the relative ineffectiveness of OISP that is revealed by Fig.~\ref{fig:gamma2}(e) originates from the proximity between the frequencies of NV$^{-}$ and electronic-like P1 transitions.

As was discussed above, when the externally applied magnetic field exceeds the value of $\sim102~\mathrm{mT}$ (i.e. above the NV$^{-}$ GSLAC) the process of OISP is expected to increase (rather than decrease) the polarization $P_{z}$. The plot in Fig.~\ref{fig:gamma2}(f) demonstrates this behavior, and shows that this process gives rise to population inversion (i.e. a change in the sign of $P_{z}$). The largest measured value of $\left| P_{z} \right|$ is about $15$ times larger
than the value corresponding to thermal equilibrium. This enhancement factor is significantly larger compared to the case
where $B<102%
%TCIMACRO{\unit{mT}}%
%BeginExpansion
\operatorname{mT}%
%EndExpansion
$ [see Fig.~\ref{fig:gamma2}(e)]. However, it is not sufficiently large to allow reaching the threshold of masing. The threshold inaccessibility occurs due to the above-discussed anomalous saturation in the dependency of polarization upon laser power.

Note that the power of microwave driving used in the CDMR measurements is about 8 orders of magnitude lower than the power used in the ODMR measurements. This fact together with the significant hyperpolarization observed in the CDMR measurements excludes resonant microwave heating effects as being responsible for the visibility of P1 transitions in the ODMR data presented above.

\section{Conclusions}

Our results demonstrate that the P1 defects in diamond and the NV$^{-}$ ground
state $+1$ to $-1$ transition can be optically probed. In addition to
electronic-like transitions in P1, we report here
for the first time on an optical detection of both nuclear-like and mixed transitions. We show
that these transitions are visible in a wide range of frequencies and
magnetic field amplitudes. These transitions may enable better sensitivity to
magnetic field, which is a key feature in quantum sensing and quantum
information. The ability to simultaneously probe the magnetic resonances of
both NV$^{-}$ and P1 can be exploited for sensitivity enhancement by using
hyperpolarization of P1 defects and exploiting the large density of P1 defects
in diamond compared to NV$^{-}$ defects ($\sim100$ times larger).

In addition, we observe an
unknown resonance feature in the ODMR spectrum. This feature is markedly clear
and is symmetric around the GSLAC magnetic field, leading us to conclude that
it corresponds to coupling between NV$^{-}$ and another unknown resonance. A significant population inversion is observed with $B>102%
%TCIMACRO{\unit{mT}}%
%BeginExpansion
\operatorname{mT}%
%EndExpansion
$. Further study is needed to understand the mechanism responsible for the anomalous saturation and to overcome it in order to reach the threshold of masing.

\section{Acknowledgements}

We greatly appreciate fruitful discussions with Paz London, Aharon Blank,
Efrat Lifshitz, Vladimir Dyakonov, Sergey Tarasenko, Victor Soltamov, Nadav
Katz, Michael Stern, Amit Finkler and Nir Bar-Gil.

\appendix

\section{Rate equations}

Consider the case where a transverse excitation having amplitude $\omega_{1}$
and angular frequency $\omega_{\mathrm{p}}$ is applied to the P1 spins having
angular transition frequency $\omega_{\mathrm{P1}}$. In addition OISP is
applied to the NV spins having angular transition frequency $\omega
_{\mathrm{NV}}$. The rate equations for the P1 and NV spin polarizations,
which are denoted by $P_{z,\mathrm{P1}}$ and $P_{z,\mathrm{NV}}$,
respectively, are taken to be given by%
\begin{align}
\frac{\mathrm{d}P_{z,\mathrm{P1}}}{\mathrm{d}t}  &  =-\frac{P_{z,\mathrm{P1}%
}-P_{z0,\mathrm{P1}}}{T_{\mathrm{P1}}}\;,\label{P_z,P1 eom}\\
\frac{\mathrm{d}P_{z,\mathrm{NV}}}{\mathrm{d}t}  &  =-\frac{P_{z,\mathrm{NV}%
}-P_{z0,\mathrm{NV}}}{T_{\mathrm{NV}}}\;. \label{P_z,NV eom}%
\end{align}
The P1 steady state value $P_{z0,\mathrm{P1}}$ is obtained by averaging the
contributions of driving-induced depolarization (relevant terms are labeled by
the subscript $\mathrm{d}$), dipolar coupling to NV spins (subscript
$\mathrm{I}$) and thermal relaxation (subscript $\mathrm{T}$)%
\begin{equation}
P_{z0,\mathrm{P1}}=\frac{T_{\mathrm{d},\mathrm{P1}}^{-1}\times0+T_{\mathrm{I}%
,\mathrm{P1}}^{-1}P_{z\mathrm{I},\mathrm{P1}}+T_{\mathrm{T},\mathrm{P1}}%
^{-1}P_{z\mathrm{T},\mathrm{P1}}}{T_{\mathrm{P1}}^{-1}}\;,
\end{equation}
where $T_{\mathrm{P1}}^{-1}=T_{\mathrm{d},\mathrm{P1}}^{-1}+T_{\mathrm{I}%
,\mathrm{P1}}^{-1}+T_{\mathrm{T},\mathrm{P1}}^{-1}$ is the total P1 rate. For
the case where the P1 transverse relaxation time $T_{2,\mathrm{P1}}$ is much
shorter than the longitudinal relaxation time the driving-induced
depolarization rate is given by%
\begin{equation}
T_{\mathrm{d},\mathrm{P1}}^{-1}=\frac{\omega_{1}^{2}T_{2,\mathrm{P1}}%
}{1+\left(  \omega_{\mathrm{p}}-\omega_{\mathrm{P1}}\right)  ^{2}%
T_{2,\mathrm{P1}}^{2}}\;.
\end{equation}
Similarly, the NV steady state value $P_{z0,\mathrm{NV}}$ is obtained by
averaging the contributions of dipolar coupling to P1 spins (subscript
$\mathrm{I}$), thermal relaxation (subscript $\mathrm{T}$) and OISP (subscript
$\mathrm{O}$)%
\begin{equation}
P_{z0,\mathrm{NV}}=\frac{T_{\mathrm{I},\mathrm{NV}}^{-1}P_{z\mathrm{I}%
,\mathrm{NV}}+T_{1\mathrm{T},\mathrm{NV}}^{-1}P_{z\mathrm{T},\mathrm{NV}%
}+T_{\mathrm{O},\mathrm{NV}}^{-1}P_{z\mathrm{O},\mathrm{NV}}}{T_{\mathrm{NV}%
}^{-1}}\;,
\end{equation}
where $T_{\mathrm{NV}}^{-1}=T_{\mathrm{I},\mathrm{NV}}^{-1}+T_{1\mathrm{T}%
,\mathrm{NV}}^{-1}+T_{\mathrm{O},\mathrm{NV}}^{-1}$ is the total NV rate.

When higher states are disregarded the P1 and NV thermal polarization
coefficients are given by $P_{z\mathrm{T},\mathrm{P1}}=-\tanh\left(
\omega_{\mathrm{P1}}/\omega_{\mathrm{T}}\right)  $ and $P_{z\mathrm{T}%
,\mathrm{NV}}=-\tanh\left(  \omega_{\mathrm{NV}}/\omega_{\mathrm{T}}\right)
$, respectively, where $\omega_{\mathrm{T}}=2k_{\mathrm{B}}T/\hbar$, $\hbar$
is the Planck's h-bar constant and $k_{\mathrm{B}}T$ is the thermal energy,
and the polarization coefficients due to dipolar coupling $P_{z\mathrm{I}%
,\mathrm{P1}}$ and $P_{z\mathrm{I},\mathrm{NV}}$ are given by%
\begin{align}
P_{z\mathrm{I},\mathrm{P1}}  &  =\tanh\left(  \frac{\omega_{\mathrm{P1}}%
}{\omega_{\mathrm{NV}}}\tanh^{-1}P_{z,\mathrm{NV}}\right)  \;,\\
P_{z\mathrm{I},\mathrm{NV}}  &  =\tanh\left(  \frac{\omega_{\mathrm{NV}}%
}{\omega_{\mathrm{P1}}}\tanh^{-1}P_{z,\mathrm{P1}}\right)  \;.
\end{align}

\newpage
\bibliographystyle{IEEEtran}
\bibliography{Eyal_Bib}
%Produces the bibliography via BibTeX.

\end{document}